\documentclass[12pt]{article}
\usepackage{graphicx}

\newcommand\pubnumber{DPF2015-259}
\newcommand\pubdate{\today}

\def\Title#1{\begin{center} {\Large #1 } \end{center}}
\def\Author#1{\begin{center}{ \sc #1} \end{center}}
\def\Address#1{\begin{center}{ \it #1} \end{center}}
\def\andauth{\begin{center}{and} \end{center}}

\newcommand\pubblock{\rightline{\begin{tabular}{l} \pubnumber\\
         \pubdate  \end{tabular}}}
\newenvironment{Abstract}{\begin{quotation}  }{\end{quotation}}
\newenvironment{Presented}{\begin{quotation} \begin{center} 
             PRESENTED AT\end{center}\bigskip 
      \begin{center}\begin{large}}{\end{large}\end{center} \end{quotation}}
\def\Acknowledgments{\bigskip  \bigskip \begin{center} \begin{large}
             \bf ACKNOWLEDGMENTS \end{large}\end{center}}

\begin{document}
\begin{titlepage}
\pubblock

\vfill
\Title{First Limits on the Dark Matter Cross-Section with the High Altitude Water Cherenkov (HAWC) Observatory}
\vfill
\Author{M. L. Proper}
\Address{Department of Physics, Colorado State University, Fort Collins, CO, USA}
\Author{J. P. Harding, B. Dingus}
\Address{Physics Division, Los Alamos National Laboratory, Los Alamos, NM, USA}
\andauth
\Author{HAWC Collaboration}
\Address{complete author list, http://www.hawc-observatory.org/collaboration/}

\vfill
\begin{Abstract}
The High Altitude Water Cherenkov (HAWC) gamma-ray observatory is a wide field-of-view observatory sensitive to $100~\rm GeV \-- 100~\rm TeV$ gamma rays and cosmic rays. The HAWC observatory is also sensitive to diverse indirect searches for dark matter annihilation, including annihilation from extended dark matter sources, the diffuse gamma-ray emission from dark matter annihilation, and gamma-ray emission from non-luminous dark matter subhalos. Among the most promising classes of objects for the indirect detection of dark matter are dwarf spheroidal galaxies. These objects are expected to have few astrophysical sources of gamma rays but high dark matter content, making them ideal candidates for an indirect dark matter detection with gamma rays. Here we present independent limits on the annihilation cross section for 14 dwarf spheroidal galaxies within the HAWC field-of-view, as well as their combined limit. These are the first limits on the annihilation cross section using data collected with HAWC.
\end{Abstract}
\vfill
\begin{Presented}
DPF 2015\\
The Meeting of the American Physical Society\\
Division of Particles and Fields\\
Ann Arbor, Michigan, August 4--8, 2015\\
\end{Presented}
\vfill
\end{titlepage}
\def\thefootnote{\fnsymbol{footnote}}
\setcounter{footnote}{0}

\section{Introduction}

Evidence for the existence of dark matter has been seen in many observations, such as in galactic rotation curves, galaxy clusters, gravitational lensing, large-scale cosmological structure, and the cosmic microwave background. While there is ample evidence of the gravitational effects of dark matter, the particle nature of dark matter remains unclear. Weakly Interacting Massive Particles (WIMPs) are among the leading hypothetical candidates for dark matter. WIMPs can annihilate into standard model particles and produce photons via pion decay, radiative processes by charged leptons, or direct production of gamma rays through loop-order processes \cite{bib:DMhawcSens}. Indirect detection of dark matter can be done by looking for the products of dark matter annihilation in objects with high dark matter content. When there is no significant detection of these by-products we can set limits on the annihilation cross section. Dwarf spheroidal galaxies are prime targets for indirect searches of dark matter as these objects are expected to have very few astrophysical sources of gamma rays, but high dark matter content. Benefiting from the low signal-to-noise ratio for these sources, we can constrain the dark matter annihilation cross section for several annihilation channels. 

\section{The HAWC Observatory}

The High Altitude Water Cherenkov (HAWC) gamma-ray observatory is a wide field-of-view observatory located at an elevation of 4100~m on the volcano Sierra Negra in the state of Puebla, Mexico. It is a second-generation water Cherenkov detector, building on the success of the Milagro experiment, and is sensitive to $100~\rm GeV \-- 100~\rm TeV$ gamma rays and cosmic rays. HAWC consists of a densely packed array of 300 Water Cherenkov Detectors (WCDs), each of which is 7.3~m in diameter and 4.5~m deep, with the total array covering an area of approximately 20,000~$\rm m^2$. Each WCD contains a light-tight hermetically sealed plastic bag, called a bladder, containing 200,000~L of ultra-purified water and is instrumented with four photomultiplier tubes (PMTs). Three of the PMTs are 8-inch Hamamatsu R5912 PMTs, re-used from the Milagro experiment, with the fourth and central PMT being a 10-inch R7081-MOD high-quantum efficient PMT. The PMTs detect Cherenkov light from energetic particles passing through the WCD, which are produced in the extensive air shower of the primary incident particle.  HAWC has an instantaneous field-of-view of 2~sr, and a duty cycle \textgreater95\%. 

\section{Gamma-Ray Emission from Annihilating Dark Matter}

\subsection{Dark matter differential flux}

A calculation of the expected gamma-ray flux from dark matter annihilation requires information about both the astrophysical properties of the potential dark matter source and the particle properties of the initial and final-state particles. The differential gamma-ray flux integrated over the solid angle of the source is: 
\begin{equation} \label{Flux}
\frac{dF}{dE} = \frac{\langle\sigma_{A}v\rangle}{8\pi M_{\chi}^2}\frac{dN_{\gamma}}{dE}J
\end{equation}
where $\langle\sigma_{A}v\rangle$ is the velocity-weighted cross section, $dN_{\gamma}/dE$ is the gamma-ray spectrum per dark matter annihilation, and $M_{\chi}$ is the dark matter mass \cite{bib:DMhawcSens}. We define the dark matter $J$-factor ($J$) as the mass density ($\rho$) squared integrated along the line-of-sight $x$, averaged over the solid angle of the observation region: 
\begin{equation} \label{Jdeltaomega}
J= \int_{\rm source} d\Omega \int dx \rho^{2} (r_{gal}(\theta,x)) 
\end{equation}
where the distance from the source is given by:
\begin{equation} \label{rgal}
r_{gal}(\theta,x) = \sqrt{R^{2} - 2xR\cos(\theta) + x^{2}}
\end{equation}
where $R$ is the distance to the center of the source, and $\theta$ is the angle between the source and the line-of-sight $x$ \cite{bib:DMhawcSens}. 

\subsection{Calculation of dark matter spectra}

Pythia 6.4 \cite{bib:Pythia} was used in this analysis to calculate the expected photon spectrum for a particular WIMP annihilation channel. The photon radiation of charged particles was simulated, as well as the decay of particles such as the $\pi^{0}$ \cite{bib:DMhawcSens, bib:ICRCpat}. For each annihilation channel and each dark matter mass, the average number of photons in each energy bin per annihilation event was calculated. This differential flux, $dN_{\gamma}/dE$, was used to determine the dark matter $J$-factor of the targeted source. The dark matter annihilation channels considered in this analysis are: $\chi\chi\rightarrow b\bar b$, $\chi\chi\rightarrow \tau^{+}\tau^{-}$, $\chi\chi\rightarrow\mu^{+}\mu^{-}$, $\chi\chi\rightarrow t\bar t$ and $\chi\chi\rightarrow W^{+}W^{-}$.

\subsection{Dark matter distributions}

In this analysis there are two dark matter density profiles that are used, the NFW (Navarro-Frenk-White) model and the Einasto model. The density profiles describe how the density ($\rho$) of a spherical system varies with distance (r) from its center. The NFW profile as developed by Julio Navarro, Carlos Frenk and Simon White \cite{bib:NFW, bib:NFW2}, is the simplest single parameter model consistent with N-body simulations. The NFW density profile takes the form: 
\begin{equation} \label{NFW}
\rho_{\rm NFW} (r)= \frac{\rho_{s}}{(r/r_{s})(1+r/r_{s})^{2}}
\end{equation}
where $\rho_{s}$ is the scale density and $r_{s}$ is the scale radius of the galaxy. The Einasto profile, as developed by Jaan Einasto \cite{bib:Einasto} is of the form:  
\begin{equation} \label{Einasto}
\rho_{\rm Einasto}(r) = \rho_{s} \exp \left[-\frac{2}{\alpha}\left(\left(\frac{r}{r_{s}}\right)^{\alpha} - 1\right)\right]
\end{equation}
where $\alpha$ controls the degree of curvature of the profile. The Einasto model is a more conservative estimate of the dark matter density profile than the NFW model.
Depending on the chosen density profile model, we can substitute $\rho(r)$ into equation \ref{Jdeltaomega} to calculate $J$ for a particular source. We use the Einasto model with $\alpha = 0.303$ for Segue1 \cite{bib:DMhawcSens}, while we use the NFW model for the remaining dwarf spheroidal galaxies. The source parameter values for the fourteen dwarf spheroidal galaxies presented in this analysis are listed in table \ref{table:sourceparameters}:
\begin{table}[htp]
\centering
{\small
\begin{tabular}{c|c|c|c|c|c|c|c|c}
\hline
\hline
Source & $RA$ & $Dec$  & $\rho_{s}$ & $r_{s}$ & $R$ & $J$ & $\sigma$ & Ref.\\ 
 & & & GeV/cm$^{3}$ & kpc & kpc & GeV$^{2}$cm$^{-5}$sr & & \\
\hline
Bootes 1 & 210.05 & 14.49 & 8.12 & 0.27 & 66 &  $3.8\times 10^{18}$ & -0.04 & \cite{bib:param1, bib:param2}\\
Canes Venatici I & 202.04 & 33.57 & 0.79 & 0.55 & 218 & $2.9\times 10^{16}$ & 0.91 & \cite{bib:param1, bib:param3} \\
Canes Venatici II & 194.29 & 34.32 & 4.77 & 0.13 & 160 &  $2.5\times 10^{16}$ & 0.34 & \cite{bib:param1, bib:param3}\\
Coma Berenices & 186.74 & 23.90 & 9.76 & 0.16 & 44 & $2.6\times 10^{18}$  & 0.88 & \cite{bib:param1, bib:DMhawcSens}\\
Draco & 260.05 & 57.07 & 0.98 & 2.1 & 76 & $2.0\times 10^{19}$ & 0.30 & \cite{bib:param1, bib:DMhawcSens}\\
Hercules & 247.72 & 12.75 & 0.80 & 0.32 & 132 & $1.6\times 10^{16}$ & -1.67 & \cite{bib:param1, bib:param3}\\
Leo I & 152.11 & 12.29 & 16.20 & 0.28 & 254 & $1.2\times 10^{18}$  & 0.13 & \cite{bib:param1, bib:param4}\\
Leo II & 168.34 & 22.13 & 162.01 & 0.06 & 233 & $1.2\times 10^{18}$ & -0.02 & \cite{bib:param1, bib:param4}\\
Leo IV & 173.21 & -0.53 & 1.99 & 0.15 & 154 & $7.3\times 10^{15}$ & 0.51 & \cite{bib:param1, bib:param3}\\
Segue 1 & 151.75 & 16.06 & 4.18 & 0.15 & 23 & $1.8\times 10^{19}$  & -0.33 & \cite{bib:param1, bib:DMhawcSens}\\
Sextans & 153.28 & -1.59 & 3.38 & 0.37 & 86 & $1.0\times 10^{18}$  & -1.55 & \cite{bib:param1, bib:param2}\\
Ursa Major I & 158.72 & 51.94 & 2.39 & 0.31 & 97 & $2.3\times 10^{17}$  & -0.37 & \cite{bib:param1, bib:param3}\\
Ursa Major II & 132.77 & 63.11 & 13.79 & 0.17 & 32 & $1.1\times 10^{19}$  & 0.10 & \cite{bib:param1, bib:param2}\\
Ursa Minor & 227.24 & 67.24 & 3.89 & 0.65 & 76 & $9.6\times 10^{18}$ & 0.26 & \cite{bib:param1, bib:param2}\\
\hline
\hline
\end{tabular}}
\caption[Source parameters]{Astrophysical parameters for the fourteen dwarf spheroidal galaxies within the HAWC field-of-view and their references. The source, right ascension ($RA$), declination ($Dec$), scale density ($\rho_{s}$) in GeV/$\rm cm^{3}$, scale radius ($r_{s}$) in kpc, distance to the source ($R$) in kpc, and the dark matter $J$-factor in GeV$^{2}$cm$^{-5}$sr are all listed above. The significance ($\sigma$) is also shown for each source. The significances listed are for $M_{\chi} = 10~\rm TeV$ and the $\chi\chi\rightarrow b \bar b$ annihilation channel. Segue 1 uses an Einasto dark matter density profile with $\alpha = 0.303$, while the remaining sources use an NFW profile.}
\label{table:sourceparameters}
\end{table}

\section{Analysis}
\subsection{Likelihood analysis}

In order to analyze a particular region of the sky, we do a likelihood ratio test. This allows us to estimate the significance of a source that has a low signal-to-noise ratio. The likelihood ratio test is a ratio of two different hypotheses: 1) the null hypothesis, which assumes that no source exists and all observed photons are due to background, and 2) the alternative hypothesis, in which a source exists. According to \cite{bib:Wilks, bib:LiMa}, we can directly relate the significance of the observed result to the likelihood ratio $L_{0}/L$, where $L_{0}$ is the null hypothesis likelihood and $L$ is the alternative hypothesis likelihood:
\begin{equation}
TS = -2\ln\left(\frac{L_{0}}{L}\right) 
\end{equation}
where $TS$ is the Test Statistic. $L_{0}$ and $L$ are maximized with respect to any free parameters. In this analysis we use data with no free parameters for $L_{0}$ and a single free parameter for $L$. Since the null and alternative hypothesis likelihoods are taken to be Poisson distributions with parameter $\mu$, then: 
\begin{equation} \label{TS}
TS  = 2 \ln\left(\frac{\mu^{k} e^{-\mu}}{k!}\right) - 2 \ln \left(\frac{\mu_{0}^{k} e^{-\mu_{0}}}{k!}\right) 
\end{equation}
For the alternative hypothesis, $\mu$ is the number of expected counts in each bin ($\mu = E+B$ where $E$ is the expected number of signal counts and $B$ is the number of background counts), and $k$ is the number of total events in each bin from data ($N$). For the null hypothesis, we make similar definitions, where $\mu_{0}$ is the number of expected counts in each bin for the null model ($B$), and $k$ is the number of total events in each bin from data ($N$). So equation \ref{TS} becomes: 
\begin{equation} \label{TS2}
TS = \sum_{bins} \left[2 N \ln\left(1+\frac{E}{B}\right)-2E\right]
\end{equation}
where $TS$ is summed over all energy-proxy bins and spatial bins \cite{bib:HAWCsens}. In this analysis the dwarf spheroidal galaxies were treated as point sources. The formalism of the likelihood ratio test has been implemented in the HAWC software, and is known as the Likelihood Fitting Framework (LiFF). The details of LiFF are described in \cite{bib:LiFF}.

\subsection{95\% confidence level limit}

We use the likelihood ratio test and definition of $TS$ in equation \ref{TS2} to constrain the dark matter annihilation cross section for a particular source. To set a 95\% Confidence Level (CL) limit, we first maximize $TS$ ($TS_{max}$) and then optimize $\Delta TS = TS_{max} - TS_{95}$ such that: 
\begin{equation}
4 = TS_{max} -\sum_{bins} \left[2 N \ln \left(1+ \frac{\xi E_{ref}}{B}\right) - 2\xi E_{ref}\right]
\end{equation}
where $\xi$ is the scale parameter with which we scale our expected counts from a source until we are 95\% confident that HAWC will be sensitive to that particular dark matter annihilation cross section. $E_{ref}$ is the expected number of counts in each bin for the reference cross section, $\langle\sigma_{A}v\rangle_{\rm ref}$. The limit is not dependent on the value chosen for $\langle\sigma_{A}v\rangle_{\rm ref}$ as our expectation is linearly proportional to the cross section, as can be seen in equation \ref{Flux}. The scale parameter $\xi$ is then used to set the 95\% CL limit for a particular source, dark matter annihilation channel and dark matter mass:%
\begin{equation} \label{limit}
\langle\sigma_{A} v\rangle_{95\%} = \xi \times \langle\sigma_{A}v\rangle_{\rm ref}
\end{equation}

\section{Dark Matter Limits from HAWC}

\begin{figure*}[t!]
\makebox[\textwidth][c]{
\begin{tabular}{cc}
\includegraphics[width=.5\textwidth]{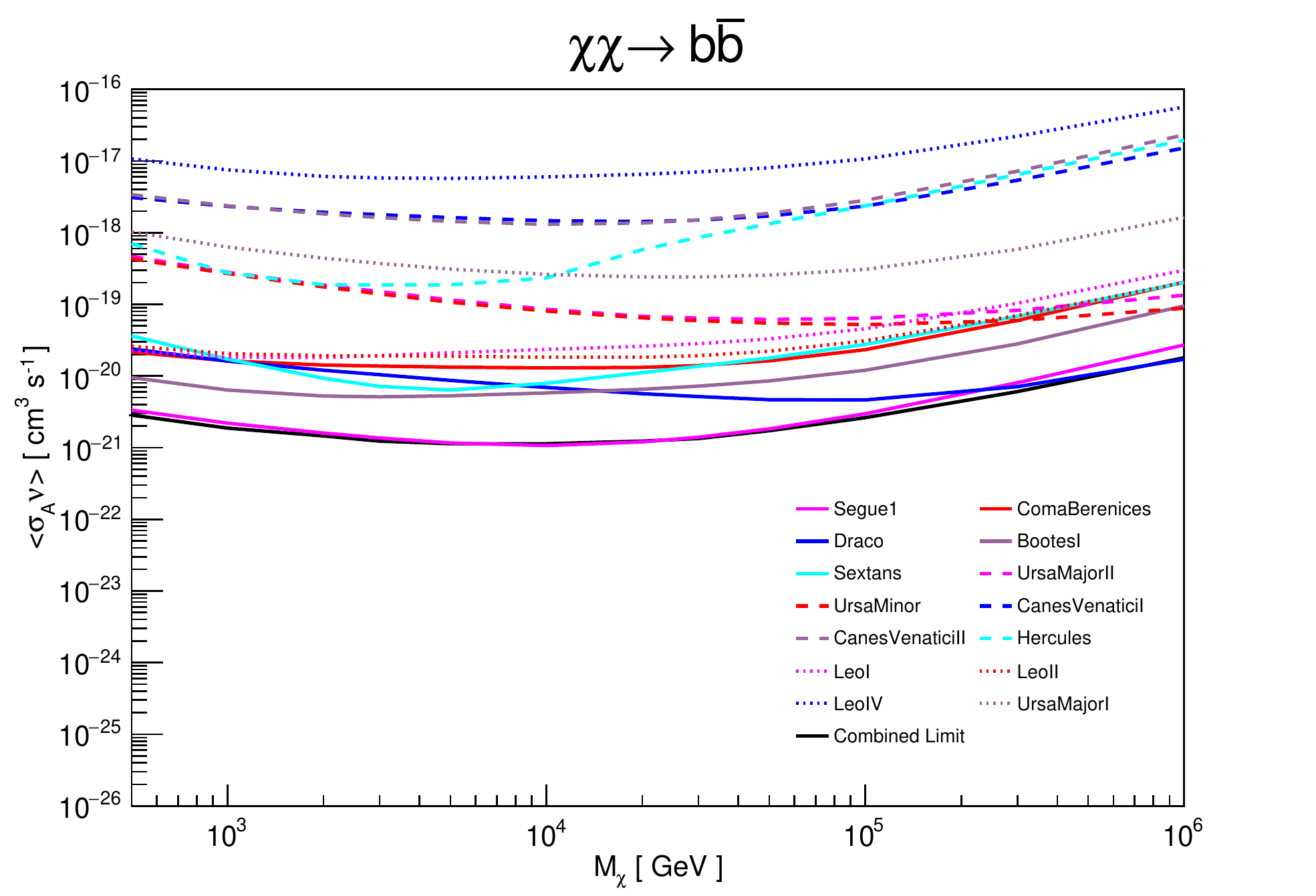}
\includegraphics[width=.5\textwidth]{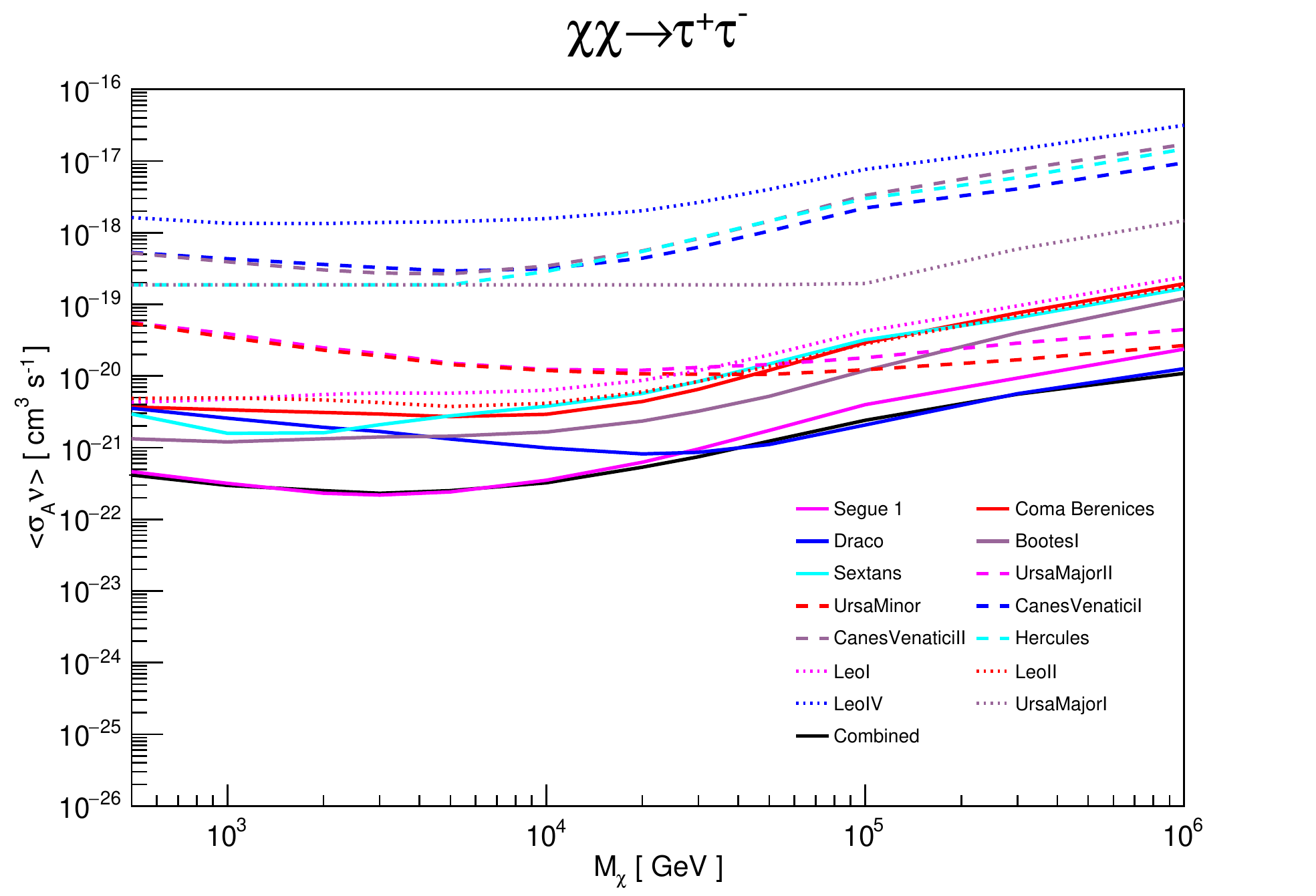}
\end{tabular}
}
\caption[caption]{\label{DMLimitsFull} The dark matter annihilation cross section limits for the fourteen dwarf spheroidal galaxies within the HAWC field-of-view. Individual limits are shown for two select dark matter annihilation channels: $\chi\chi\rightarrow b\bar b$ and $\chi\chi\rightarrow \tau^{+}\tau^{-}$. The solid black line is the combined limit from a stacked analysis of all fourteen dwarf spheroidal galaxies for the shown annihilation channels.}
\end{figure*}

\begin{figure*}[tp!]
\makebox[\textwidth][c]{
\begin{tabular}{cc}
\includegraphics[width=.5\textwidth]{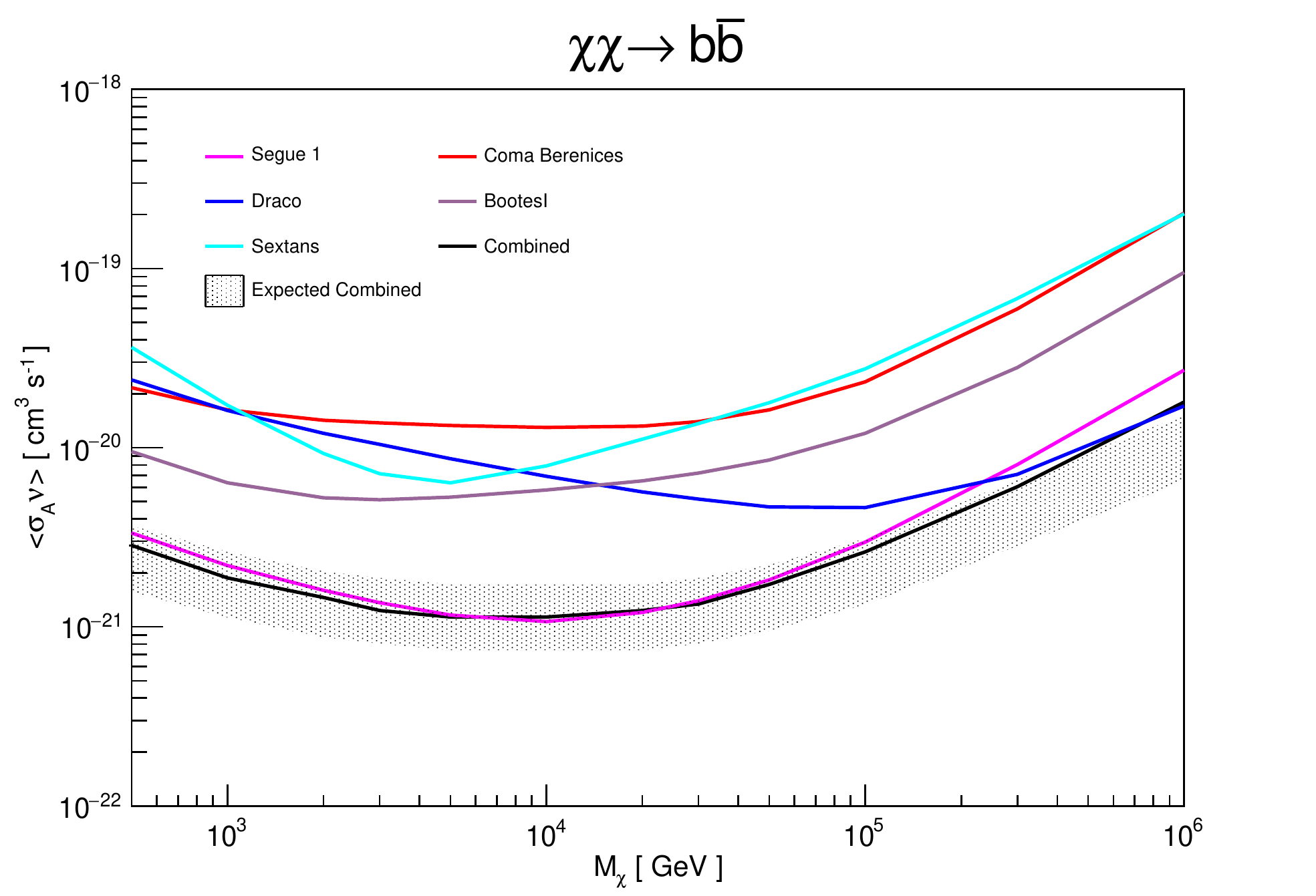}
\includegraphics[width=.5\textwidth]{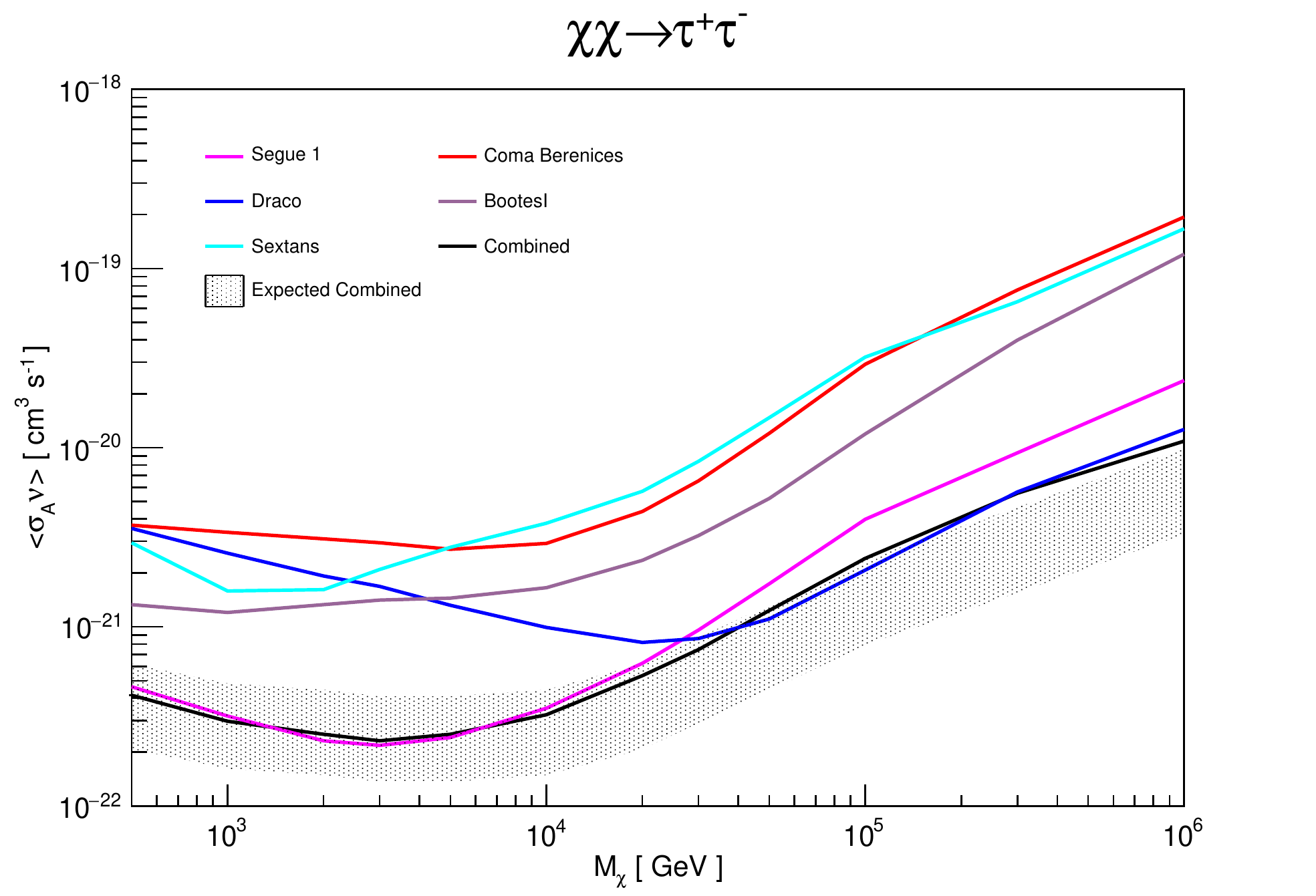}\\
\includegraphics[width=.5\textwidth]{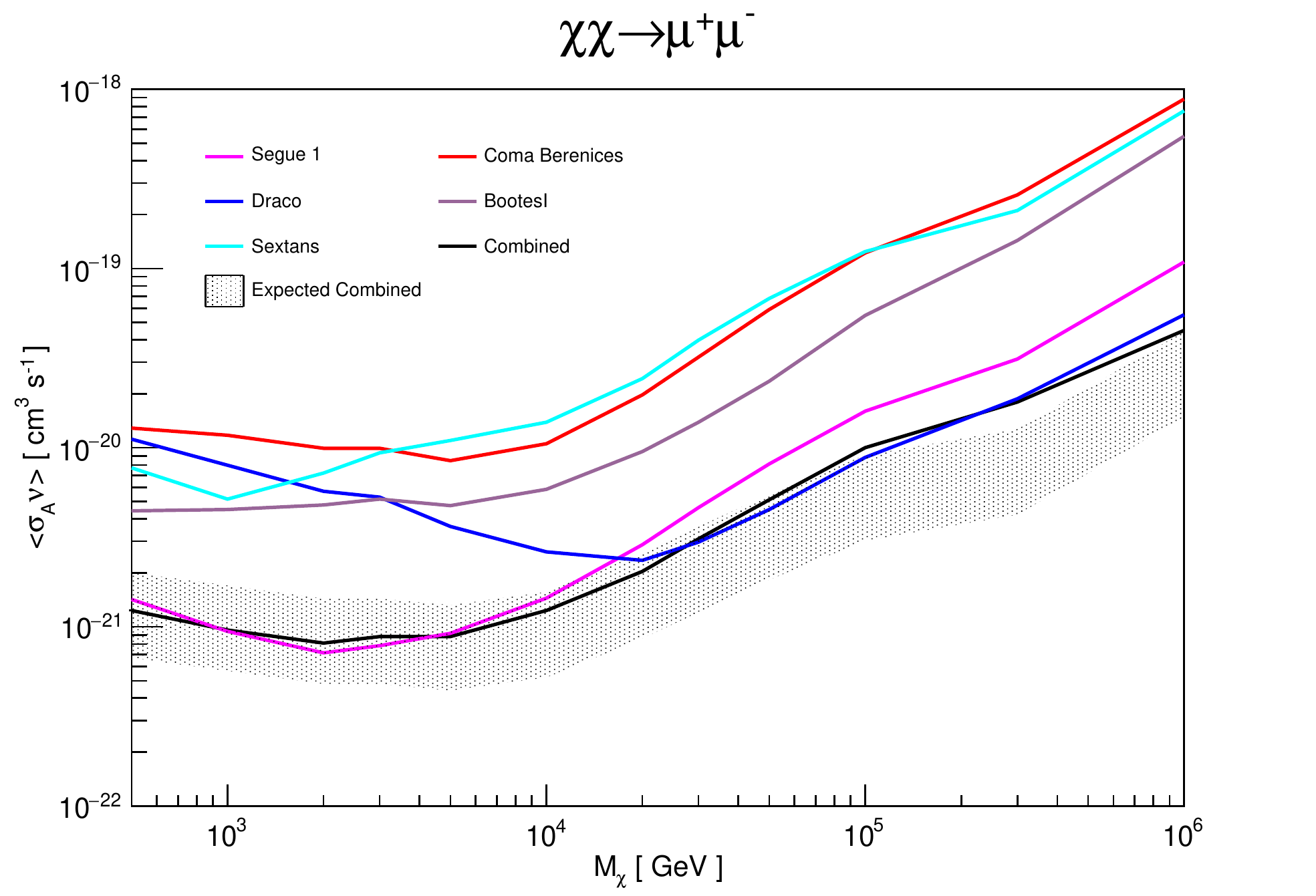}
\includegraphics[width=.5\textwidth]{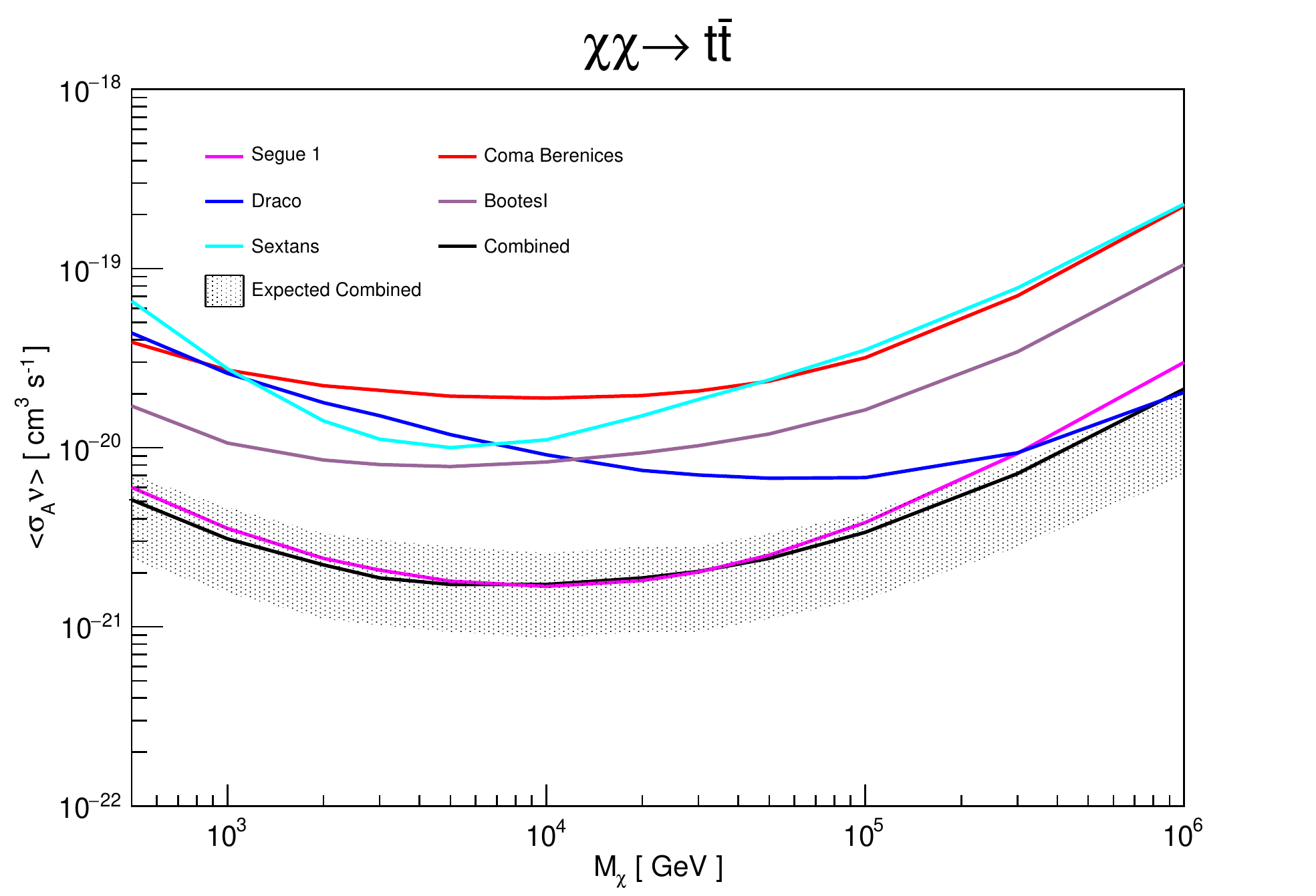}\\
\includegraphics[width=.5\textwidth]{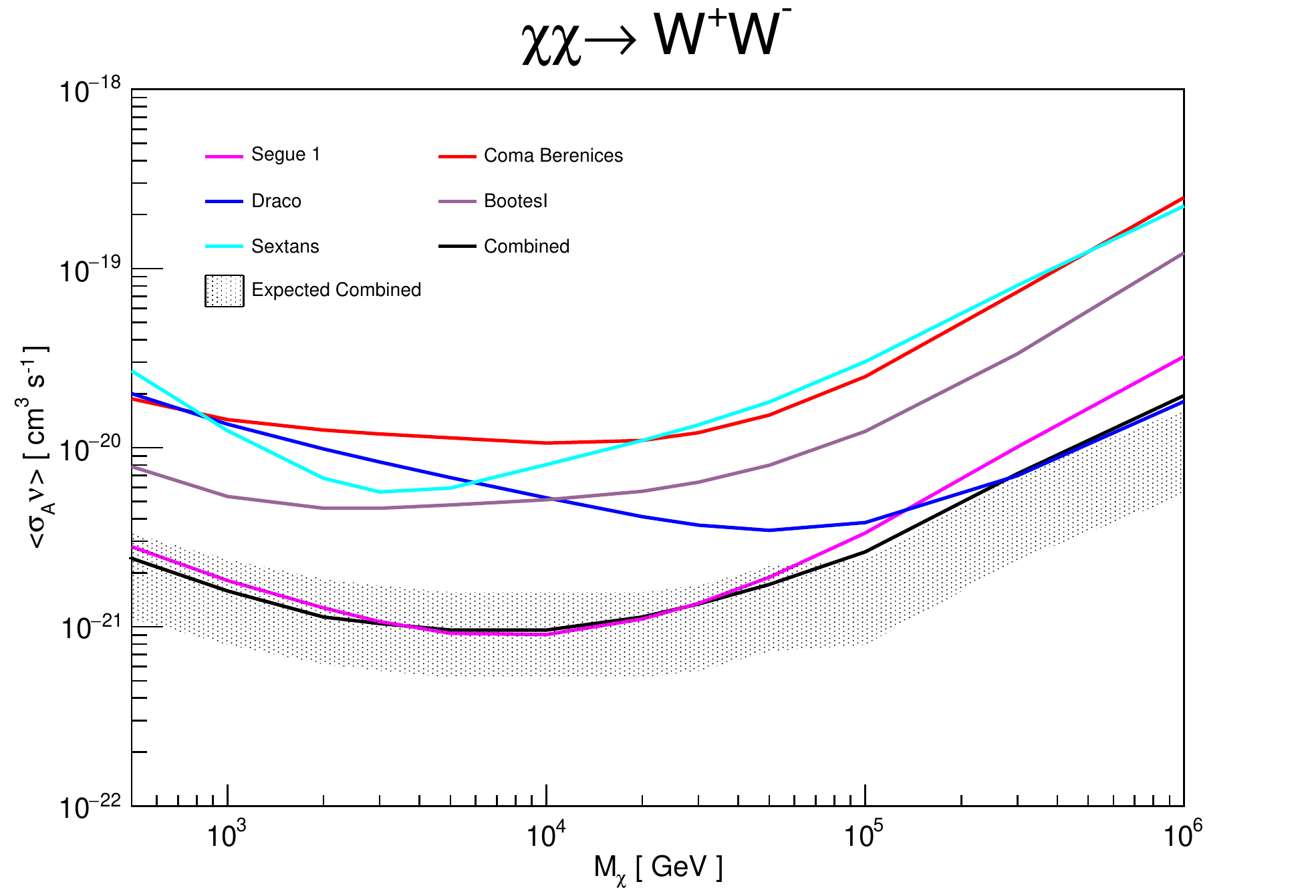}
\includegraphics[width=.5\textwidth]{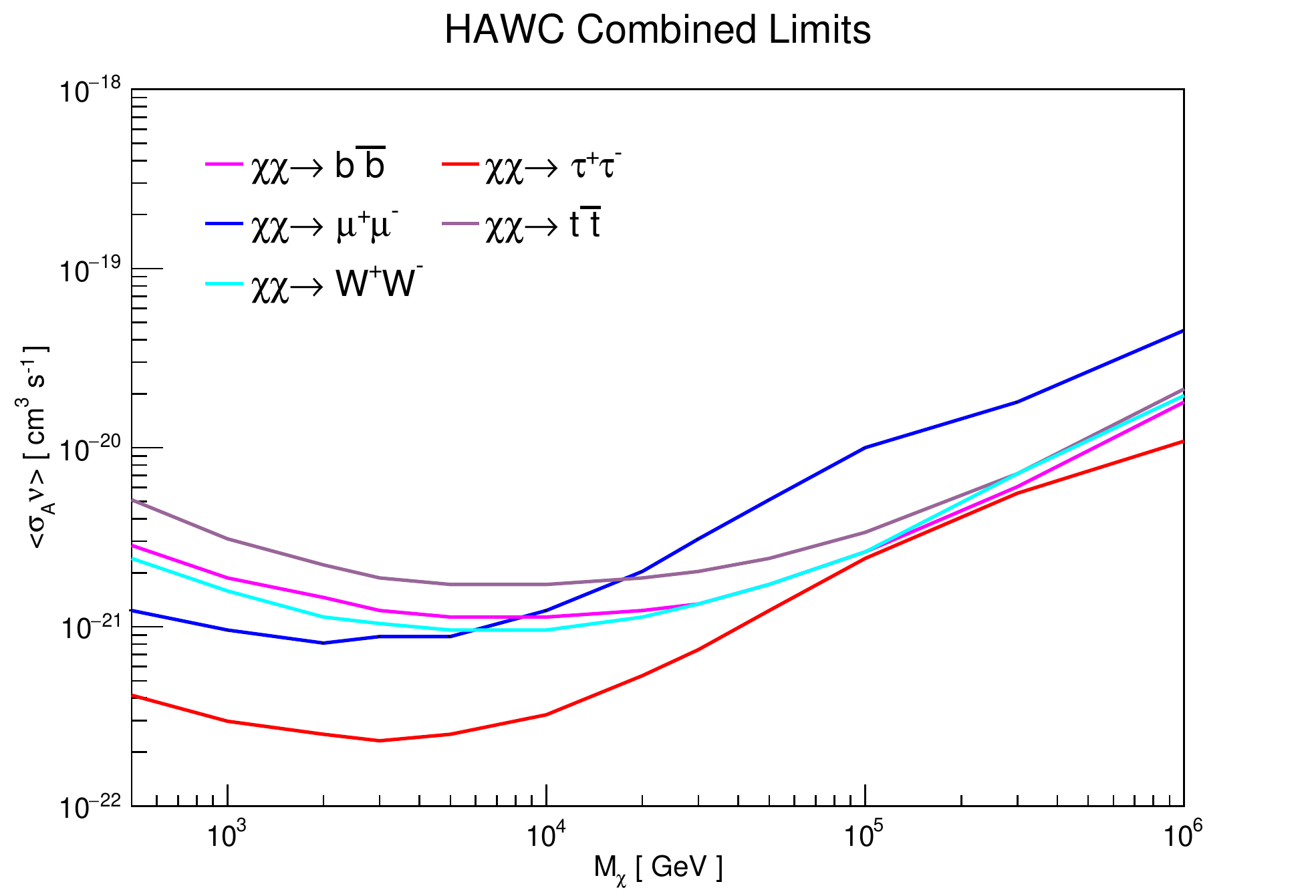}
\end{tabular}
}
\caption[caption]{\label{DMLimits} Individual dark matter annihilation cross section limits for the five best dwarf spheroidal galaxies within the HAWC field-of-view: Segue 1, Draco, Sextans, Coma Berenices and BootesI. The solid black line in the first five figures shows the combined limit for each annihilation channel as a result of a stacked analysis with all fourteen dwarf spheroidal galaxies, while the hatched areas show the expected fourteen source combined limits with systematic error on the expected signal. The lower right figure shows the five annihilation channel combined limits together.}
\end{figure*}

We present in this analysis individual limits from the fourteen dwarf spheroidal galaxies within the HAWC field-of-view. The limits shown here are done assuming the dwarf spheroidal galaxies act like point sources, not extended sources. Through detailed simulation of the HAWC gamma-ray sensitivity and backgrounds \cite{bib:HAWCsens}, we have determined the significance of the gamma-ray flux for a range of dark matter masses, $0.5~\rm TeV \-- 1000~\rm TeV$, and five annihilation channels: $\chi\chi\rightarrow b\bar b$, $\chi\chi\rightarrow \tau^{+}\tau^{-}$, $\chi\chi\rightarrow\mu^{+}\mu^{-}$, $\chi\chi\rightarrow t\bar t$ and $\chi\chi\rightarrow W^{+}W^{-}$. We can then convert the source significance into exclusion curves of the dark matter annihilation cross section for the individual dwarf spheroidal galaxies, as detailed in section 4. The data used in this analysis was taken while the array was under construction between August 2, 2013 and July 8, 2014, during which the active detector sites grew from 362 PMTs in 108 WCDs to 491 PMTs in 134 WCDs.  This analysis considered WIMPs annihilating with 100\% branching ratios into $b\bar b$, $\tau^{+}\tau^{-}$, $\mu^{+}\mu^{-}$, $t\bar t$ and $W^{+}W^{-}$. 

The 95 \% CL limits for the $b\bar b$ and $\tau^{+}\tau^{-}$ annihilation channels, for the fourteen dwarf spheroidal galaxies within the HAWC field-of-view, are shown in figure \ref{DMLimitsFull}. Figure \ref{DMLimits} shows the exclusion curves for five annihilation channels, for the dwarf spheroidal galaxies that HAWC is most sensitive to: Segue 1, Draco, Sextans, Coma Berenices and BootesI. A combined limit was done for each annihilation channel by doing a stacked analysis for the fourteen dwarf spheroidal galaxies, and by optimizing the total expected counts from all sources. The combined limits are presented both in figures \ref{DMLimitsFull} and \ref{DMLimits}. The HAWC expected fourteen source combined limit was also calculated for each annihilation channel and is shown in figure \ref{DMLimits}, with the hatched areas showing the systematic uncertainty on the expected signal from simulation as discussed in \cite{bib:Errors}. 

\section{Summary}
In these proceedings we present individual 95\% CL limits on the annihilation cross section for fourteen dwarf spheroidal galaxies within the HAWC field-of-view. A combined limit is also shown from a stacked analysis of all dwarf spheroidal galaxies. Limits are presented for a range of dark matter masses and for several dark matter annihilation channels, resulting from data collected over a 180 day period during construction of the HAWC detector. These are the first limits on the dark matter annihilation cross section presented with HAWC data. Further analysis of this work will include treating the dwarf spheroidal galaxies as extended sources, as well as producing exclusion curves for data collected with the entire HAWC array, where the limits are expected to improve by about an order of magnitude.

\Acknowledgments
%
We acknowledge the support from: the US National Science Foundation (NSF);
the US Department of Energy Office of High-Energy Physics;
the Laboratory Directed Research and Development (LDRD) program of
Los Alamos National Laboratory; Consejo Nacional de Ciencia y Tecnolog\'{\i}a (CONACyT),
Mexico (grants 260378, 55155, 105666, 122331, 132197, 167281, 167733);
Red de F\'{\i}sica de Altas Energ\'{\i}as, Mexico;
DGAPA-UNAM (grants IG100414-3, IN108713,  IN121309, IN115409, IN111315);
VIEP-BUAP (grant 161-EXC-2011);
the University of Wisconsin Alumni Research Foundation;
the Institute of Geophysics, Planetary Physics, and Signatures at Los Alamos National Laboratory;
the Luc Binette Foundation UNAM Postdoctoral Fellowship program.

\end{document}